\documentclass[letterpaper]{article}
\usepackage{aaai2026}  
\usepackage{times}  
\usepackage{helvet}  
\usepackage{courier}  
\usepackage[hyphens]{url}  
\usepackage{mwe} 
\usepackage{graphicx} 
\usepackage{natbib}  
\usepackage{caption} 
\usepackage{algorithm}
\usepackage{booktabs}
\usepackage{algorithmic}
\usepackage{multirow}
\usepackage{subfig}

\usepackage{newfloat}
\usepackage{listings}
\usepackage{amsmath} 
\DeclareCaptionStyle{ruled}{labelfont=normalfont,labelsep=colon,strut=off} 
\floatstyle{ruled}
\newfloat{listing}{tb}{lst}{}
\floatname{listing}{Listing}
\newcommand{\abr}[1]{\textsc{#1}}
\usepackage{bibentry}
\usepackage{xcolor}
\definecolor{lavender}{RGB}{230, 230, 250}
\usepackage{framed}       
\usepackage{changepage}   
\usepackage{enumitem}     

\definecolor{formalshade}{rgb}{0.8, 0.9, 0.8}
\newenvironment{formal}{%
  \MakeFramed{\advance\hsize-\width\FrameRestore}%
  \noindent\hspace{-4.55pt}
  \begin{adjustwidth}{}{7pt}%
  \vspace{0.5pt}\vspace{0.5pt}%
}
{%
  \vspace{0.5pt}\end{adjustwidth}\endMakeFramed%
}

\usepackage{fvextra}
\DefineVerbatimEnvironment{WrappedVerbatim}{Verbatim}{
    breaklines=true
}

\title{RescueLens: LLM-Powered Triage and Action on Volunteer Feedback for Food Rescue}
\author {
    Naveen Raman,\textsuperscript{\rm 1} 
    Jingwu Tang,\textsuperscript{\rm 1}
    Zhiyu Chen,\textsuperscript{\rm 2}
    Zheyuan Ryan Shi,\textsuperscript{\rm 3}
    Sean Hudson,\textsuperscript{\rm 4}
    Ameesh Kapoor,\textsuperscript{\rm 4}
    Fei Fang\textsuperscript{\rm 1}
}
\affiliations {
    \textsuperscript{\rm 1} Carnegie Mellon University\\
    \textsuperscript{\rm 2} University of Texas at Dallas\\
    \textsuperscript{\rm 3} University of Pittsburgh\\
    \textsuperscript{\rm 4} 412 Food Rescue\\

}

\begin{document}
\newcommand{\nrcomment}[1]{{\color{red} Naveen: {#1}}}
\newcommand{\zhiyu}[1]{\textcolor{red}{[zhiyu: #1]}}
\newcommand{\ryan}[1]{\textcolor{blue}{[ryan: #1]}}

\maketitle

\begin{abstract}
Food rescue organizations simultaneously tackle food insecurity and waste by working with volunteers to redistribute food from donors who have excess to recipients who need it. 
Volunteer feedback allows food rescue organizations to identify issues early and ensure volunteer satisfaction. 
However, food rescue organizations monitor feedback manually, which can be cumbersome and labor-intensive, making it difficult to prioritize which issues are most important. 
In this work, we investigate how large language models (LLMs) assist food rescue organizers in understanding and taking action based on volunteer experiences. 
We work with 412 Food Rescue, a large food rescue organization based in Pittsburgh, Pennsylvania \footnote{\url{https://412foodrescue.org}}, to design \textbf{\abr{RescueLens}}, an LLM-powered tool that automatically categorizes volunteer feedback, suggests donors and recipients to follow up with, and updates volunteer directions based on feedback.
We evaluate the performance of \abr{RescueLens} on an annotated dataset, and show that it can recover 96\% of volunteer issues at 71\% precision.
Moreover, by ranking donors and recipients according to their rates of volunteer issues, \abr{RescueLens} allows organizers to focus on 0.5\% of donors responsible for more than 30\% of volunteer issues. 
\abr{RescueLens} is now deployed at 412 Food Rescue and through semi-structured interviews with organizers, we find that \abr{RescueLens} streamlines the feedback process so organizers better allocate their time. 
\end{abstract}

\section{Introduction}
\label{sec:intro}
Despite advances in food production, 800 million people remain chronically undernourished worldwide~\citep{unicef_food_security}. 
At the same time, 40\% of the food produced worldwide is wasted, demonstrating that food insecurity is a problem of distribution rather than production~\citep{driven_to_waste,food_wastage_footprint}. 
Food rescue organizations simultaneously tackle both food waste and insecurity by redistributing food from those who have excess to those in need.
Food rescue organizations work with volunteers to organize rescue trips, where volunteers transport food from donors, such as grocery stores, to recipients, including shelters. 
Food rescue organizations have been massively successful, saving over  150 million pounds of food in the United States since 2016~\citep{food_rescue_hero}. 

Food rescue organizations rely on volunteer feedback to understand the issues faced by volunteers. 
Volunteer feedback is critical because it serves as the primary mechanism for volunteers to learn about volunteer behaviors.
Without it, these organizations have little visibility into relational issues between volunteers, donors, and recipients. 
For example, volunteer feedback can alert food rescue organizations to situations where a grocery store repeatedly misses their food pickup or if volunteers have repeated issues dropping off food for a particular recipient. 
Organizers at food rescue organizations can then intervene by contacting donors, changing pickup schedules, or updating directions. 

While volunteer feedback is important, going from feedback to action is difficult. 
Although only a small fraction of volunteers leave feedback, this quickly adds up. 
For example, 412 Food Rescue, a large food rescue organization in Pittsburgh, received more than 1800 pieces of feedback per year, which costs the organization time. 
Moreover, manual feedback tracking makes it difficult to determine how best to allocate organizer time because it is unclear which issues are the most pressing or occur most frequently.  

In this work, we investigate how large language models (LLMs) can help food rescue organizations understand volunteer feedback.  
LLM-based tools are typically faster and more efficient than human analysts, which can free up organizer time~\citep{generative_ai_copilot_impact}. 
For example, LLM-based tools can quickly alert organizers to situations when intervention is needed, allowing organizers to focus on the most pressing pieces of feedback. 
At the same time, food rescue organizations are typically resource-limited, limiting their ability to develop large datasets for LLM applications. 
Moreover, volunteer feedback can often be ambiguous and involve domain-specific context. 

\begin{figure*}
    \centering 
    \includegraphics[width=\textwidth]{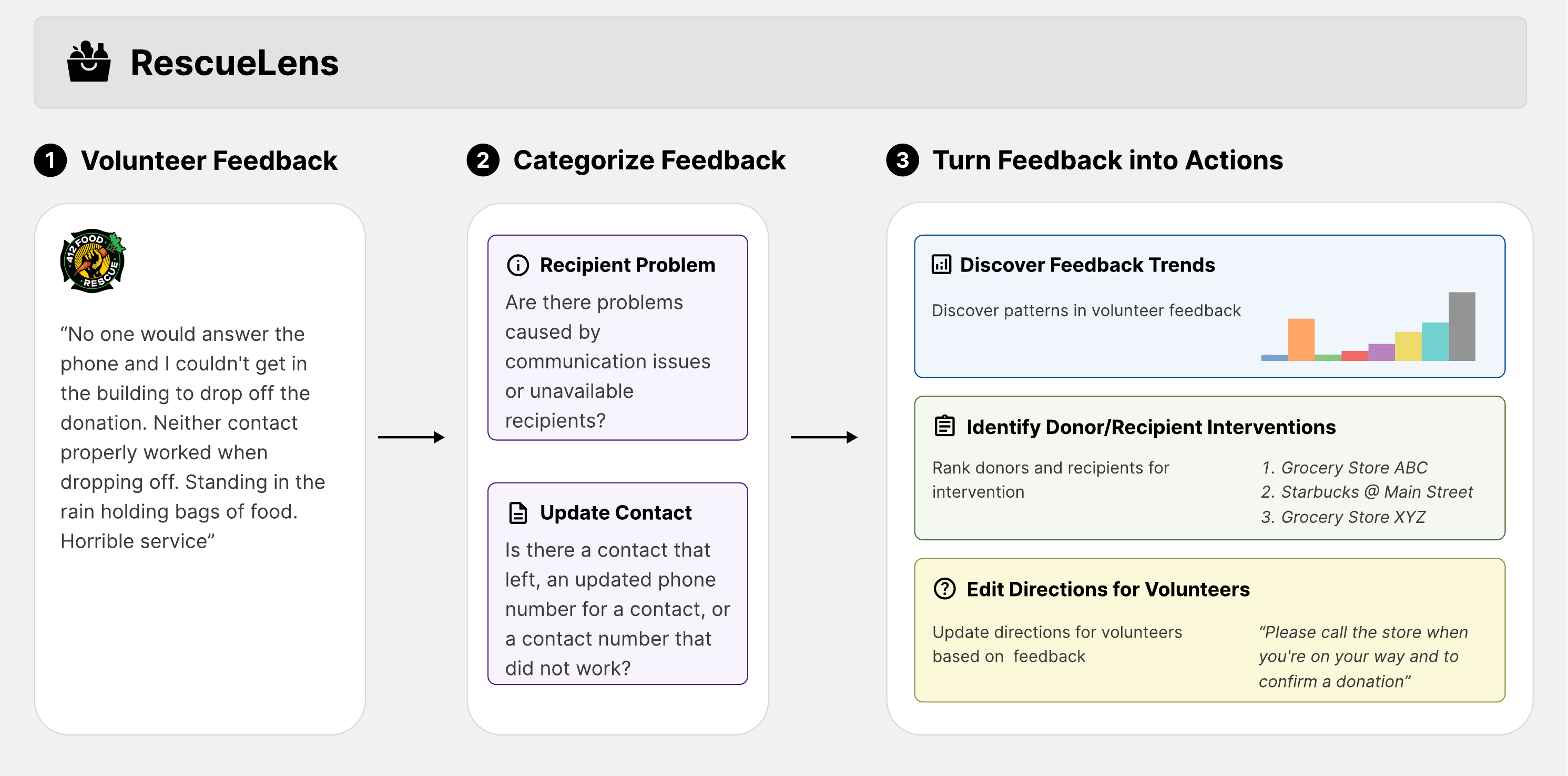}
    \caption{We introduce \abr{RescueLens}, an LLM-powered tool that automatically analyzes volunteer feedback in food rescue.
    Our tool first categorizes volunteer issues into different categories, such as \colorbox{lavender}{Recipient Problem} and \colorbox{lavender}{Update Contact}. \abr{RescueLens} then uses these predictions to discover trends in volunteer feedback, identify which donors and recipients require interventions, and suggest updates to the directions based on feedback.}
    \label{fig:pull}
\end{figure*}

In response to the challenges faced by food rescue organizations, we develop \textbf{\abr{RescueLens}}, an LLM-powered tool that automatically analyzes feedback and enables organizers to take actions based on these insights (see Figure~\ref{fig:pull} for a summary).
We built \abr{RescueLens} in coordination with 412 Food Rescue, a large food rescue organization based in Pittsburgh, Pennsylvania, that has rescued over 30 million pounds of food since 2015. 
We first conducted a need-finding study with organizers at 412 Food Rescue, where we found that there is currently little formal documentation of volunteer feedback analyses, making it difficult to track which issues are most pressing.
Based on this user study, we designed \abr{RescueLens} to consist of two components: 1) an LLM-based system which uses LLM with few-shot learning to efficiently categorize feedback and 2) a set of action modules that leverage feedback categorizations to a) identify which donors and recipients require intervention and b) update volunteer directions based on feedback. 
Through a mixed methods study, we demonstrate that \abr{RescueLens} achieves a 96\% recall and 71\% precision, while allowing organizers to focus on the 0.5\% of donors responsible for more than 30\% of volunteer issues. 
Through interviews with organizers, we show that \abr{RescueLens} helps quantify which problems are most pressing and determine how best to allocate their time. 
Our tool has been deployed at 412 Food Rescue since May 2025, and has analyzed more than 1,200 pieces of feedback from volunteers so far.
Beyond food rescue, \abr{RescueLens} can be broadly applied across non-profits to better understand their text-based feedback. 
\footnote{We include our code in the attached supplementary material, and we will release it publicly after the camera-ready period. Furthermore, some of our code is in the 412 Food Rescue repository, which is private}
\section{Related Work}
\label{sec:related}
\paragraph{Food Rescue}
Computational work in food rescue can be categorized into algorithmic research, which analyzes matching algorithms between volunteers, donors, and recipients, and system-level research, which investigates platform design~\citep{ai_for_social_good}. 
Food rescue organizations leverage algorithms both for matching donors with recipients~\citep{automating_food_drop} and notifying volunteers about rescue trips~\citep{global_rewards,food_rescue_recommender_system,context_rmab}. 
Matching is difficult because organizers balance volunteer engagement, allocation efficiency, and fairness~\citep{fair_division_food,global_rewards,branch_and_bound_food_rescue}. 
On the other hand, system-level research investigates how to improve system design by analyzing volunteer interactions.
Examples include one work that deploys a notification system at a university to reduce food waste~\cite{pittgrub} and another that informs users about task difficulty~\citep{task_difficulty}.
Our work can be viewed as an extension of~\citet{task_difficulty}, where we investigate how to design tools that mitigate task difficulties. 

\paragraph{LLMs and Non-Profits}
Beyond food rescue, our work is broadly situated in the field of using LLMs to assist non-profits. 
Developing LLMs in non-profit scenarios is more challenging due to limitations in computational power and dataset size~\citep{challenges_non_profit}. 
In our situation, this required us to use in-context learning rather than fine-tuning. 
Additionally, developing LLMs for non-profits requires a balance between an LLM's abilities and its risks for hallucinations and bias~\citep{llm_non_profit}. 
Moreover, non-profits themselves might have varied perspectives or desiderata for using LLMs. 
For example, some non-profits are concerned with the governance of LLMs, while others are concerned with the environmental impact of LLMs~\citep{non_profit_governance}.
These diverse perspectives inspire us to conduct interviews with organizers to understand \textit{how} organizers planned to use \abr{RescueLens} and tailor modules. 
\paragraph{Automatic Analysis of User Feedback}
Non-profit stakeholders often express subjective feedback that can help improve systems, but automatic approaches are necessary to gain insights from large-scale data~\citep{ai_traffic_simulator,decision_language_model}. 
Subjective stakeholder feedback is prevalent across domains, including  education~\citep{education_feedback}, e-commerce~\citep{ecommerce_feedback}, and mobile applications~\citep{app_feedback}. 
While stakeholder feedback is prevalent, automatic analysis is hard due to domain-specific language and ambiguity~\citep{education_feedback_analysis_trends}. 
To circumvent this, prior work categorized feedback into different topics through supervised methods such as BERT and LLMs \citep{llm_feature_analysis} and unsupervised methods such as topic modeling~\citep{international_student_topic_model}. 
Our work extends these feedback analysis techniques to scenarios with little labeled training data, then combines them with tools that transform user feedback into actionable insights for organizers.
\section{Motivating RescueLens: A Needfinding Study at 412 Food Rescue}
\label{sec:user}
\paragraph{Study Procedure}
To better understand current practices for processing volunteer feedback, we conducted a series of needfinding studies with three organizers at 412 Food Rescue.
Volunteer attrition is a large issue in food rescue organizations~\citep{task_difficulty}.
A better understanding of volunteer feedback allows organizers to combat volunteer attrition by understanding volunteer problems.
To understand current practices around volunteer feedback, we recruited three organizers (P1, P2, and P3) from 412 Food Rescue through social media and email advertisements. 
We asked each organizer a series of questions related to their role at 412 Food Rescue, current practices for processing volunteer feedback, and any issues that organizers face  (we include all questions in Appendix~\ref{sec:questions_needfinding}).
We then discuss a potential tool that automatically analyzes volunteer feedback and ask organizers for feedback. 
We received approval from our institution's IRB, and we compensated  \$30 for the 30-minute study.   

\paragraph{User Study Results}
Our user study revealed that the current feedback analysis procedure is largely manual with little formal documentation. 
Because the process lacks formal documentation, organizers mention difficulties in understanding which issues are most pressing and which only occur rarely. 
For example, P2 notes ``We can't keep track of [feedback] as much as we need to.''
Such issues are amplified due to the scale of food rescue operations, with P2 noting that ``800 recipients and 500 donors are too much to keep up with.''

When we ask organizers to envision the benefits an automated system could potentially have, we find that organizers value tools that identify problems and fix volunteer directions based on feedback. 
P1 notes that ``It's helpful to have a clear-cut list of issues you may experience with the app.''
P3 similarly notes that ``coding these issues is helpful because so many different people make touch points'', revealing the utility in categorizing feedback.
Organizers also detail the utility of having tools for identifying which donors and recipients require intervention, with P3 stating ``It would be useful to have a database that points out where they need to work and what problems we need to focus on.''
Organizers also note that having a system automatically edit directions would be useful, with P2 noting that ``It would be useful if we could get pinged for the [updated] directions.''

\section{RescueLens System Design}
\label{sec:architecture}
\abr{RescueLens} is an LLM-based system that first categorizes volunteer feedback, then uses these classifications to recommend actions for food rescue organizers. 

\subsection{Categorizing User Feedback}
\subsubsection{Volunteer Dataset}
\label{sec:dataset}
\begin{figure}[t]
    \centering 
    \includegraphics[width=0.2\textwidth]{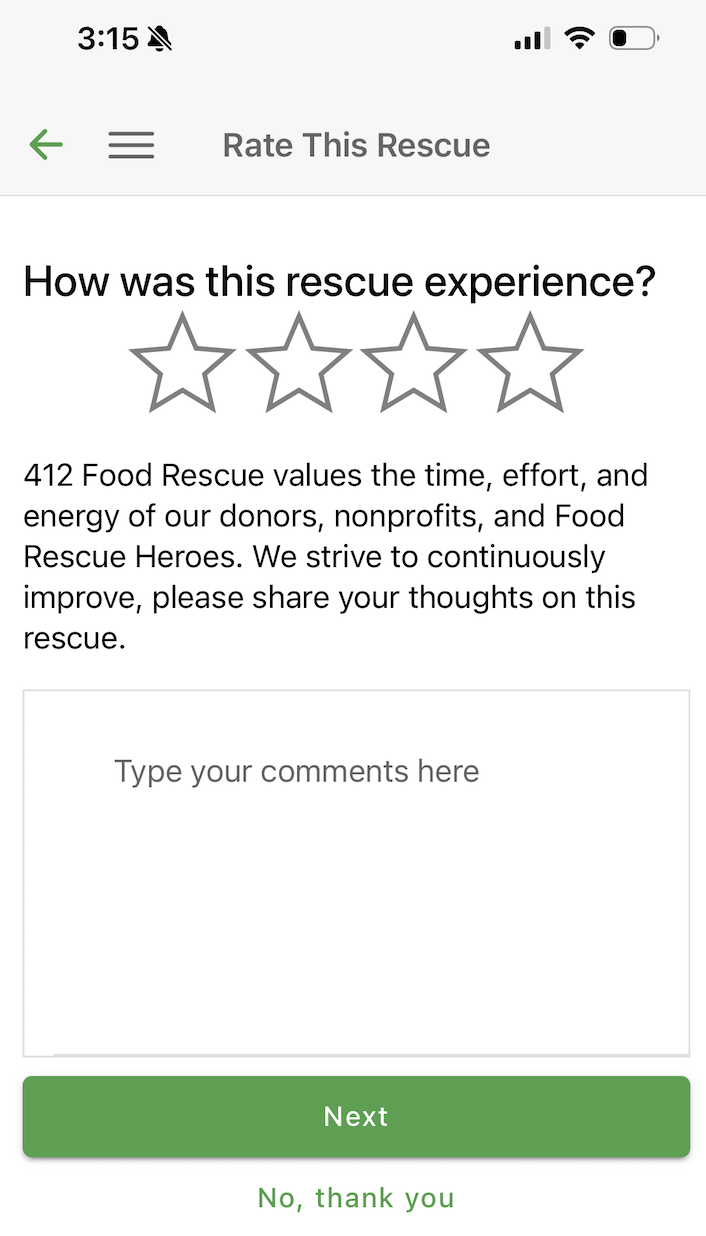}
    \hspace{1em} 
    \includegraphics[width=0.2\textwidth]{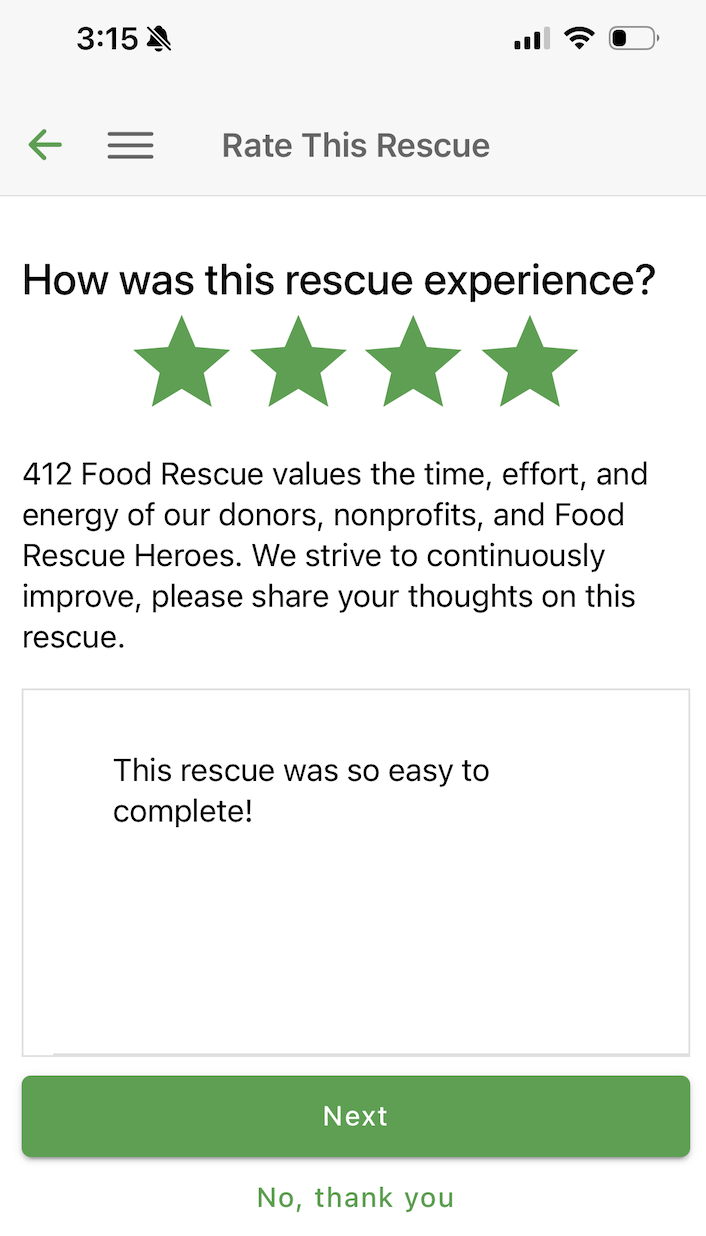}

    \caption{After each rescue trip, 412 Food Rescue collects a rating out of four along with text-based feedback.}
    \label{fig:feedback_screenshot}
\end{figure}

To construct \abr{RescueLens}, we begin with a dataset of volunteer feedback from rescue trips. 
412 Food Rescue elicits user feedback after each rescue trip along with a rating out of four (see Figure~\ref{fig:feedback_screenshot}).
The dataset starts from 2018 and includes over 200000 rescue trips. 
In total, this results in 14,439 pieces of text feedback in the database. 

\subsubsection{Defining User Feedback Categories}
\label{sec:categories}
As a first step towards automatic feedback analysis, we perform open coding to understand the types of feedback present. 
We start with a random sample of 250 volunteer feedback from rescue trips in 2024, then employed open coding, iteratively refining the codebook as new themes emerged. 
We stopped coding after observing exhaustion, where no new categories arose. 
After discarding categories that did not influence organizer action (e.g., comments that gave positive reviews of their trip), we arrived at seven categories, which we define below: 
\begin{enumerate}
    \item \colorbox{lavender}{Inadequate Food}: Assess whether the reported challenges or failures in the food rescue process were caused by inadequate food quantities provided by the donor. For example, ``\textit{Nothing to donate. Everything they had put aside was burned.}''
    \item \colorbox{lavender}{Earlier Pickup}: Assess whether the reported challenges or failures in the food rescue process were caused by someone else (e.g., another volunteer) picking up the food earlier, leading to little or no available food. For example, ``\textit{Per the store manager, someone else was already there today and picked up everything.}'' 
    \item \colorbox{lavender}{Donor Problem}: Assess whether the reported challenges or failures in the food rescue process were caused by communication issues with donors. For example, ``\textit{Terrible pickup!  Nobody knew who 412 was. After 1/2 hour, I was given 3 boxes of apples. As I left, I was flagged down and given a cart full of leftover Easter candy.}''
    \item \colorbox{lavender}{Recipient Problem}: Assess whether communication issues or unavailable recipients caused the reported challenges. For example, ``Food pantry was closed.''
    \item \colorbox{lavender}{Update Contact}: Assess whether the feedback discussed the need to update contact information for a donor or recipient. For example, ``\textit{The manager contact at Walmart has a new job and won't be there starting next week.}''
    \item \colorbox{lavender}{System Problem}: Assess whether bugs or glitches on the food rescue app caused the reported challenges. For example, ``\textit{System not responsive.}''
    \item \colorbox{lavender}{Direction Problem}: Assess whether unclear or inaccurate directions or information caused the reported challenges.  For example, ``\textit{The map directions took me to Alexander Street. Please adjust pick up location to Powell.}'' 
\end{enumerate}

\subsubsection{Employing LLMs for Classification}
\label{sec:predicting}
We construct prompts for each category. 
Prompts consist of background data, general task description, specific task details, and a few-shot examples with manually written explanations.
We use in-context learning, which improves LLM performance without the need for large annotated corpora~\citep{brown2020language}.
Each task has 3-8 few-shot examples, and we include details on all prompts in  Appendix~\ref{sec:prompts}. 

\subsection{Turning Feedback into Actions}
\abr{RescueLens} converts feedback classifications into actionable insights through a pair of tools: the first ranks donors and recipients for intervention, and the second rewrites volunteer directions. 

\subsubsection{Determining Where to Intervene}
\label{sec:donor_recipient}
We built a module in \abr{RescueLens} to help food rescue organizers identify which donors and recipients require attention.
The module produces a ranked list of donors and recipients, each scored using two types of feedback: volunteer ratings and reported issues.
The score represents the rate at which volunteers have issues when completing rescue trips for a donor or recipient. 
Higher scores represent higher priorities for intervention.  
We use these scores to rank donors and recipients, which can help organizers decide where to focus their efforts. 

We compute the score based on the volunteer rating score, on a 1-4 scale, and a set of predictions from \abr{RescueLens}.
From the predictions made by \abr{RescueLens}, we produce a comment score that represents whether any issue is present. 
For donors, we compute whether \colorbox{lavender}{Update Contact}, \colorbox{lavender}{Inadequate Food}, \colorbox{lavender}{Earlier Pickup}, \colorbox{lavender}{Direction Problem}, or \colorbox{lavender}{Donor Problem} is present, while for recipients, we check whether \colorbox{lavender}{Update Contact} or \colorbox{lavender}{Direction Problem} or \colorbox{lavender}{Recipient Problem} is present.
The final score is then the fraction of rescue trips where either the comment score is non-zero or the rating score is below 4. 

\subsubsection{Suggesting Direction Rewrites}
We constructed a tool to automatically rewrite volunteer directions for feedback labeled as \colorbox{lavender}{Direction Problem}.
Each donor or recipient has a set of directions that includes information on driving directions, points of contact, and delivery details. 
While directions are critical to volunteer success, they can become outdated over time. 
To assist with this, we periodically process the latest batch of new feedback and, for each feedback item, use an LLM to rewrite the relevant volunteer directions.
For each feedback item, we first prompt the LLM to determine whether it contains new information that warrants updating the directions; only if the LLM identifies relevant updates do we generate a revised direction.
We prompt LLMs with the original directions, the new feedback, and explicit constraints to incorporate all new information (e.g., updated contact, entrance, or address), preserve correct existing details, and remove information only if directly contradicted. 
We use seven manually curated few-shot examples—covering both donor and recipient cases—to guide the rewrite style and scope. 
We include details on our prompts and few-shot examples in Appendix~\ref{sec:prompts}. 
\section{RescueLens Evaluation}
\label{sec:experiments}
We evaluate the performance of \abr{RescueLens} through comparisons with baselines on historical data. 

\subsection{Evaluation Setup}
\label{sec:setup}
We evaluate the classification module of \abr{RescueLens} by comparing with baselines on expert-annotated data. 
We randomly sample 125 data points consisting of volunteer feedback from 2024, and then we annotate feedback for each category from Section~\ref{sec:categories}. 
We use two annotators, having them each independently rate metrics, before coming to consensus. 
We assess performance according to four metrics: accuracy, precision, recall, and F1 score.
We measure the inter-annotator agreement through Cohen's $\kappa$, and find that it is $\kappa=0.73$, indicating significant agreement across independent annotators.

We compare \abr{RescueLens} against LLM and non-LLM baselines.
For LLM baselines, we maintain the prompts used by \abr{RescueLens} while varying the underlying LLM.
By default, \abr{RescueLens} uses GPT-4o mini~\citep{gpt4o}, and we compare this against GPT-4o~\citep{gpt4o}, Llama 3~\citep{llama3}, and DeepSeekR1~\citep{deepseekr1}. 
We additionally compare \abr{RescueLens} against a term frequency-inverse document frequency (TF-IDF) baseline with logistic regression, which computes the relative frequency of different words, and a DistilBERT module~\citep{distilbert}, which finetunes BERT using a small dataset of labeled volunteer feedback. 
All evaluations are across three random seeds. 

\subsection{Classification Evaluation}
\label{sec:model}
\begin{table}[]
\centering 
\caption{We evaluate \abr{RescueLens} on an annotated dataset. The first four rows are LLM-based approaches, while the latter two are non-LLM approaches. We find that \abr{RescueLens}, which relies on a GPT-4o mini backbone, outperforms all alternatives on accuracy and F1 score except GPT-4o. Moreover, \abr{RescueLens} is significantly cheaper than both GPT-4o and DeepSeek R1, demonstrating that \abr{RescueLens} is both cost-efficient and well-performing.}
\footnotesize 
\begin{tabular}{@{}lrrrrr@{}}
\toprule
Model       & Acc. & Prec. & Recall & F1 & Cost \\ 
\midrule
\textbf{RescueLens} & \multicolumn{5}{c}{} \\
\quad +GPT-4o mini &     93.1\%                         &              83.3\%                 &    97.4\%    &  89.8\%    & \$15.00     \\
\quad +GPT-4o      &      94.4\%                         &              88.1\%                 &    94.9\%    &  91.4\%  &  \$250.00    \\
\quad +Llama 3.1   &       90.5\%                         &              76.5\%                 &    100.0\%    &  86.7\% & \$10.00 \\
\quad +DeepSeek R1 &      79.4\%                         &              68.6\%                 &    61.5\%    &  64.9\%   & \$80.00   \\ 
\cmidrule(lr){1-6}
TF-IDF      &  31.0\%                         &              31.0\%                 &    100.0\%    &  47.3\%   & \multicolumn{1}{c}{-}      \\
DistilBERT  &  31.0\%                         &              31.0\%                 &    100.0\%    &  47.3\% & \multicolumn{1}{c}{-}   \\ \bottomrule

\end{tabular}
\label{tab:eval}

\end{table}

\paragraph{Comparison against Baselines}
In Table~\ref{tab:eval}, we compare the performance and cost per year when varying the model. 
We compare the rates of finding any issue; that is, the rate of detecting whether any of the seven categories are true, given a comment. 
We focus on this metric because it represents the success of \abr{RescueLens} on finding which comments require further organizer intervention. 
We show that GPT-4o maximizes accuracy and F1 score, while GPT-4o mini sacrifices a bit of performance for cost reduction. 
\abr{RescueLens}.
Moreover, LLM-based algorithms outperform non-LLM baselines; all LLM-based variants of \abr{RescueLens} achieve at least 75\% accuracy, while TF-IDF and DistillBERT achieve 31\% accuracy, as they always predict true. 
Comparing between LLMs, we find that \abr{RescueLens}, which is built on top of GPT-4o mini, is 2\% worse in F1 score compared to GPT-4o, while reducing costs from \$250 to \$15 per year. 
While Llama 3.1 costs \$5 less than GPT-4o mini, it performs worse on the F1 score, demonstrating that GPT-4o mini is Pareto efficient among these models. 

\begin{figure}[h]
    \centering 
    \includegraphics[width=0.45\textwidth]{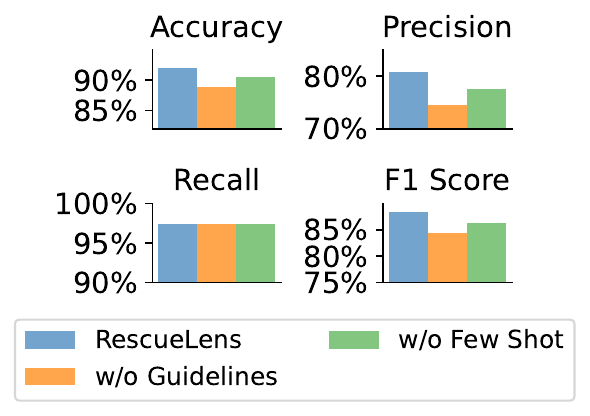}
    \caption{We conduct an ablation study to understand the importance of different aspects of \abr{RescueLens}. We find that \abr{RescueLens} performs well because of a combination of few-shot learning and task-specific guidelines.}
    \label{fig:ablation}
\end{figure}

\paragraph{Ablating RescueLens Components}
\label{sec:ablation}

To understand which components are most responsible for the performance of \abr{RescueLens}, we ablate different components. 
The prompt for \abr{RescueLens} consists of a description of the task and scenario, a set of detailed guidelines stating what to label or not label, and a set of few-shot examples with explanations (see Appendix~\ref{sec:prompts} for examples). 
For example, one part of the guidelines for \colorbox{lavender}{Donor Problem} states to ``mark comments where the interaction with the donor was delayed as donor problems.''
We ablate the guidelines and few-shot examples, then re-evaluate the performance of \abr{RescueLens} after this. 
In Figure~\ref{fig:ablation}, we compare the model performance across these three variants, and find that removing the guidelines reduces F1 score by 4\%, while removing few-shot learning reduces the F1 score by 2\%.  
Most of this reduction is concentrated in a lower precision, as precision reduces by 6\% and 3\% when removing guidelines and few-shot examples, respectively. 

\paragraph{Performance by Category}
\begin{table}[]
\centering 
\caption{We evaluate the performance of \abr{RescueLens} across different categories. We find that \abr{RescueLens} achieves at least a 75\% recall across categories, and achieves over 93\% accuracy across all categories.}
\small 
\begin{tabular}{@{}lrrrr@{}}
\toprule
Category       & Acc. & Prec. & Recall & F1 \\ 
\midrule
Any Issue   &  93.1\%                      &              83.3\%              &    97.4\%    &  89.8\% \\
\quad Donor Problems   &  95.2\%                      &              68.9\%                &    100.0\%   &  81.5\%  \\
\quad \quad Inadequate Food      &  94.2\%                       &              73.3\%                &    100.0\%    &  84.6\%    \\
\quad \quad Earlier Pickup      &  100.0\%                        &            100.0\%                &   100.0\%    &  100.0\%     \\
\quad \quad Donor Problem      & 93.7\%                       &              42.5\%                &    83.3\%    &  56.0\%   \\
\quad Recipient Problem      & 96.6\%                        &              48.1\%              &    100.0\%   &  65.0\%     \\
\quad   Update Contact    &  97.9\%                       &              27.8\%               &    100.0\%    &  43.3\%    \\
\quad   System Problem    &  98.4\%                      &             75.0\%               &    75.0\%   & 75.0\%     \\
\quad  Direction Problem     &  98.7\%                         &             75.4\%               &    100.0\%   &  85.9\%   \\
\bottomrule

\end{tabular}
\label{tab:category}
\end{table}

We stratify the performance of \abr{RescueLens} by each of the seven feedback categories from Section~\ref{sec:categories}.
In Table~\ref{tab:category}, we find that \abr{RescueLens} achieves at least a 75\% recall rate across categories, and at least a 92\% accuracy score as well. 
We intentionally design \abr{RescueLens} to optimize for recall rather than precision after internal discussions because organizers can filter down false positives.
High recall ensures that organizers are able to find all volunteer comments that require further action. 
Categories like \colorbox{lavender}{Earlier Pickup} and \colorbox{lavender}{Inadequate Food} have high rates for recall and precision because they are less ambiguous, and these categories also have the highest inter-annotator agreements. 
While \abr{RescueLens} has a low precision for \colorbox{lavender}{Donor Problem}, we note that this occurs because of mix-ups within other donor-related issues, as shown through the higher score across all donor problems. 
Across all categories, \abr{RescueLens} achieves a higher recall than precision; we intentionally construct \abr{RescueLens} in such a way that organizers can pare down false positives when intervening. 
Moreover, for some categories, such as the low F1 score, it is due to mislabeling between different types of donor issues, as shown by the higher F1 score across all donor problems. 

\subsection{Evaluating Direction Rewrites}
We evaluate the performance of the direction rewrite module against a set of three criteria: 
\begin{enumerate}
    \item \textbf{Helpfulness} - Does the new direction contain information that is important for completing the trip? For example, does it contain information on directions, contact information, or important logistics? 
    \item \textbf{Novelty} - How well does the new direction incorporate information from both the original direction and the feedback? Does it properly incorporate the new information, without removing any important previous information? 
    \item \textbf{Clarity} - How clear are the directions written; can they be easily understood without much thought? 
\end{enumerate}
We have two authors annotate 30 rewritten directions. 
Each direction is assessed on a 1-5 scale for each criterion, and we write out the scoring rubric in Appendix~\ref{sec:rubric}. 
Our inter-annotator agreement is $\kappa=0.39$, indicating fair agreement.

\begin{figure}[t]
    \centering 
    \includegraphics[width=0.475\textwidth]{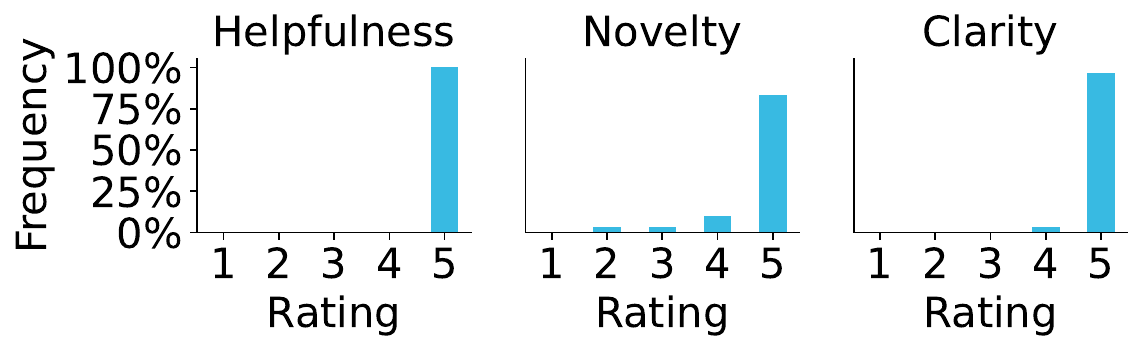}

    \caption{We assess the performance of our direction rewrite module according to three criteria: helpfulness, novelty, and clarity. Across these criteria, we find that \abr{RescueLens} performs well, averaging over a 4.7/5 across all three categories.}
    \label{fig:direction_rewrite}
\end{figure}

In Figure~\ref{fig:direction_rewrite}, we plot the distribution of scores across each of the three criteria after averaging scores between annotators.. 
We find that \abr{RescueLens} performs well across categories; for all three categories, the average score is at least 4.7/5. 
Moreover, over 70\% of the rewritten sentences achieved a perfect score of 5/5 across all three categories, demonstrating that \abr{RescueLens} can construct rewritten directions that are helpful, novel, and clear. 
Rewritten directions are almost always clear and can incorporate new volunteer feedback most of the time. 
\section{Deploying RescueLens}
\label{sec:deployment}
We provide details on our deployment of \abr{RescueLens} to 412 Food Rescue, then perform a mixed methods study to evaluate the impact of \abr{RescueLens}. 

\subsection{Deployment to 412 Food Rescue}
\label{sec:deployment_info}
We deployed \abr{RescueLens} at 412 Food Rescue through a series of stages where different versions of \abr{RescueLens} were deployed. 
We began development of \abr{RescueLens} in Spring 2024 and produced the first working prototype early in Fall 2024. 
We spent the next several months integrating \abr{RescueLens} into 412 Food Rescue, and we produced our first report on February 11th, 2025. 
We worked on further updates to better integrate \abr{RescueLens}, and officially launched our \abr{RescueLens} on May 15th, 2025, where it analyzed over 600 pieces of feedback so far. 

We integrate \abr{RescueLens} into the existing database for 412 Food Rescue through a new table titled ``Rescue Feedback.'' 
This table includes information on each piece of feedback, its categorization into seven different categories, and notes left by organizers during analysis. 
By integrating directly into the database, we ensure easy access and usage by organizers at 412 Food Rescue. 
To automatically populate this table, we run a daily Ruby script that calls \abr{RescueLens} to populate the database with the previous day's feedback. 
Each feedback is labeled daily with the seven categories from Section~\ref{sec:categories}. 
To incorporate our action modules into 412 Food Rescue, we send the direction rewrites and the ranked list of donors and recipients on a monthly basis.
We update the action modules monthly because the donor and recipient rankings are aggregations of historic comments that operate over longer timescales, while direction rewrites occur only a few times per week. 

During deployment, we made changes to \abr{RescueLens} to improve usability. 
We introduce a user interface that allows organizers to search historical data.  
Organizers can query the user interface across date ranges and combinations of volunteer feedback classifications.
The user interface also allows organizers to track which pieces of feedback require intervention and to leave notes.
We also present information from \abr{RescueLens} to organizers via an automatic bot in Slack that fetches a daily list of analyzed rescues. 

\subsection{Impact on Practice}
\label{sec:deployment_eval}
We assess the impact of \abr{RescueLens} through a mixed methods study. 
We first demonstrate how \abr{RescueLens} reduces organizer workload by guiding their actions towards intervening on the most impactful donors and recipients. 
We complement this with a semi-structured interview with an organizer at 412 Food Rescue, who discusses the impact of \abr{RescueLens} in reducing the organizer's workload and streamlining the feedback analysis process. 

\begin{figure}[t]
    \centering 
    \includegraphics[width=0.45\textwidth]{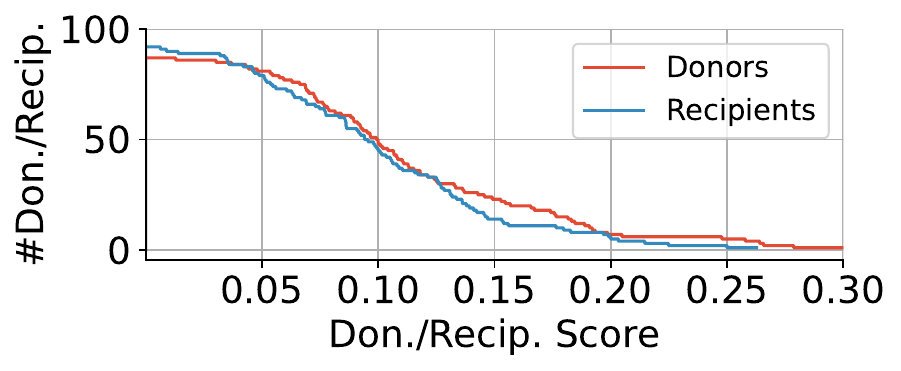}

    \caption{We compute scores for donors and recipients using the formula from Section~\ref{sec:donor_recipient}, then plot the distribution of scores. We show that only a few donors and recipients require intervention, and by focusing on these few, we can reduce organizer efforts.}
    \label{fig:donor_survival}
\end{figure}

\begin{figure}[t]
    \centering 
    \includegraphics[width=0.45\textwidth]{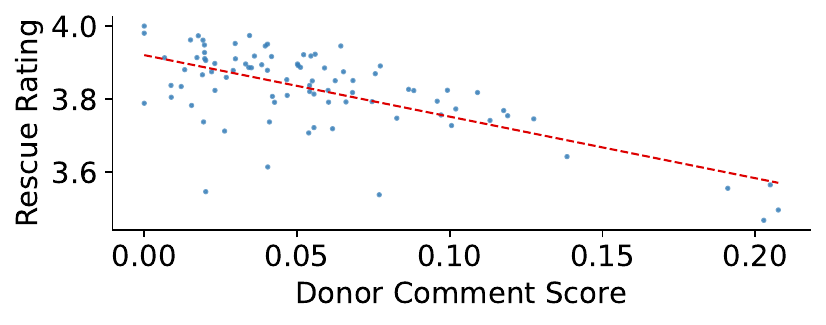}
    \includegraphics[width=0.45\textwidth]{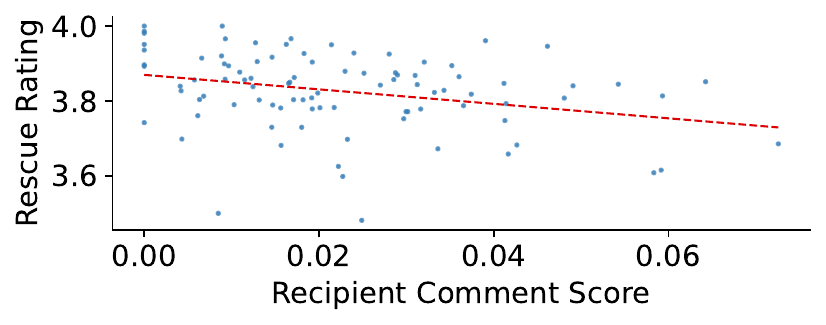}

    \caption{We plot the correlation between donor and recipient comment scores and the average rating for rescue trips associated with that donor or recipient. We find a large negative correlation ($r^{2} = 0.45$) for donor comment scores and a moderate correlation for recipient comment scores ($r^{2} = 0.09$). Fixing donor-related issues can potentially improve overall volunteer experiences.}
    \label{fig:donor_correlation}
\end{figure}

\begin{figure*}[t]
    \centering 
    \includegraphics[width=\textwidth]{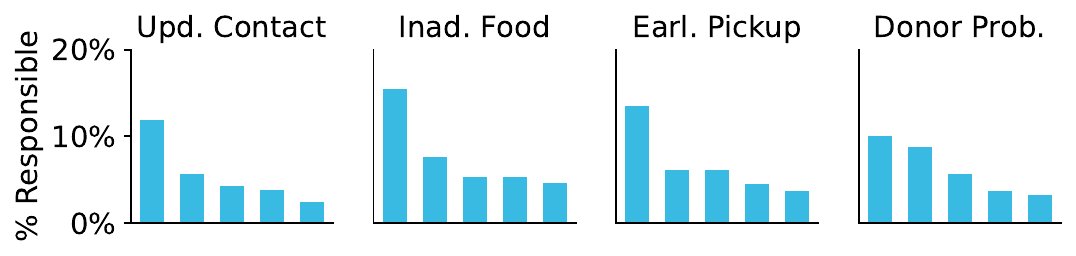}

    \caption{Here, we plot the five donors with the largest number of comments for each of four categories. Although there are hundreds of donors, we show that a small subset of donors is responsible for many of the issues. By directing organizers towards these donors, we can reduce the organizer workload by focusing on the donors who require intervention.}
    \label{fig:donor_distro}
\end{figure*}

\abr{RescueLens} ranks donors and recipients, which enables organizers to focus on the donors and recipients who lead to the most volunteer issues. 
In Figure~\ref{fig:donor_survival}, we compute the distribution of scores for donors and recipients with at least 100 rescue trips. 
Here, we use the same score from Section~\ref{sec:donor_recipient}, where a higher score indicates a higher rate of volunteer issues with the donor or recipient. 
We find that the vast majority of donors and recipients have a low score, indicating positive volunteer experiences, and less than 20\% of donors and recipients have scores larger than $0.20$. 
Addressing these issues can improve volunteer experiences with 412 Food Rescue; in Figure~\ref{fig:donor_correlation}, we show that rescue trip ratings correlate with a donor's comment score ($r^{2} = 0.45$) and somewhat correlate with a recipient's comment score ($r^{2} = 0.09$). 
Moreover, addressing these issues only requires intervention on a few donors. 
In Figure~\ref{fig:donor_distro}, we show that 5 donors, representing 0.5\% of all donors, are responsible for at least 25\% of all comment issues (across each category) for each of the 4 categories. 
Moreover, these donors have an issue rate disproportionate to their size; they only cover 2.5\% of rescue trips but are responsible for 25\% of issues. 
We stratify this trend by donor issue in Figure~\ref{fig:donor_distro} and show that five donors, representing $0.5\%$ of all donors, are responsible for more than 30\% of volunteer issues across four categories. 
By providing organizers with a ranked list of donors and recipients for intervention, we guide organizers towards donors most responsible for issues.

\begin{figure}[t]
    \centering 
    \includegraphics[width=0.48\textwidth]{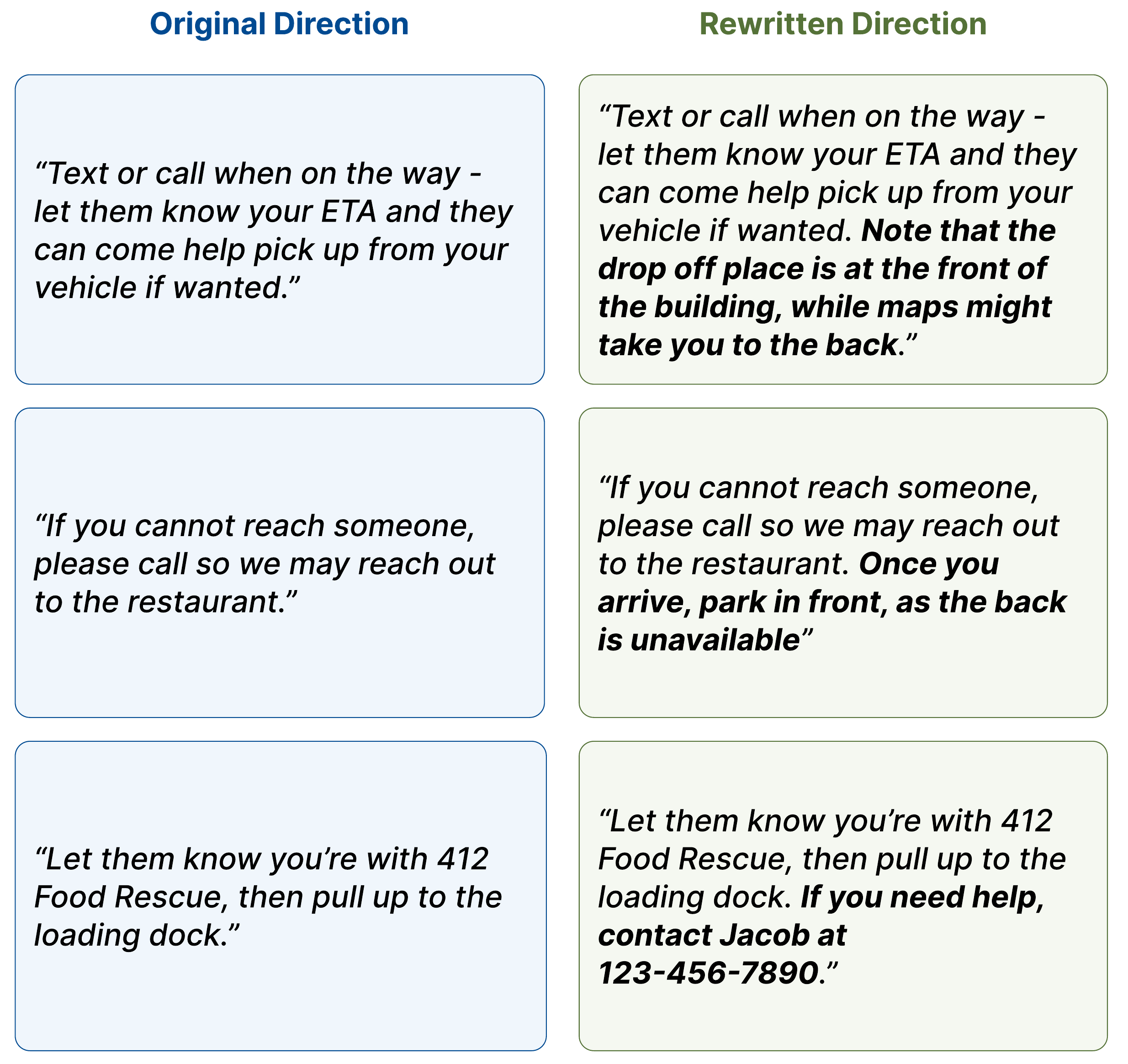}

    \caption{We plot examples of rewritten directions. These directions incorporate valuable information on contact information and drop-off details.}
    \label{fig:rewritten_directions}
\end{figure}

We additionally find that they can include valuable information based on volunteer feedback. We show three examples of rewritten directions in Figure~\ref{fig:rewritten_directions}.
Each direction includes vital information, such as phone numbers, drop-off details, and driving directions. 
Including these rewritten directions helps improve volunteer experiences by making it clearer how to complete each rescue. 
Such instructions can also help volunteers avoid common pitfalls. 

To understand the impact of \abr{RescueLens} in practice, we conducted a semi-structured interview with organizers at 412 Food Rescue (we include all questions in Appendix~\ref{sec:questions_needfinding}).
We conducted a 30-minute interview with the donor relations coordinator at 412 Food Rescue. 
The donor relations coordinator handles reach-outs to donors based on volunteer feedback, making them a natural fit to assess the impact of \abr{RescueLens}. 
We asked a series of questions related to the impact and performance of RescueLens in practice. 

Through the interview, we find that \abr{RescueLens} reduces the manual effort needed to understand user feedback while streamlining the outreach process. 
The organizer reported on the daily impact of \abr{RescueLens}:
\begin{formal}
    We look at \abr{RescueLens} every day so we can track everyday trends with food donors, such as shifts in store leads or changes in ordering. 
\end{formal}
\abr{RescueLens} is helpful because organizers are able to understand volunteer patterns and turn these into actions: 
\begin{formal}
    Because of \abr{RescueLens}, we are able to see where the trends are happening seasonally or quarterly, and it helps us quantify and qualify what needs to be fixed. 
\end{formal}
\abr{RescueLens} enables organizers to take actions based on volunteer feedback, as the organizer noted: 
\begin{formal}
    Around 50\% of the reported comments from \abr{RescueLens} are actionable, and we try to get to every actionable thing. 
\end{formal}
Our interview demonstrates the power of \abr{RescueLens} in enabling organizers at 412 Food Rescue to better understand volunteer feedback and take action based on this. 
\section{Lessons Learned}
\label{sec:lessons}
Our experience deploying \abr{RescueLens} has provided a set of valuable insights, which we detail below:  
\begin{enumerate}
    \item \textbf{Importance of Integration and Presentation} - Much of the work in developing \abr{RescueLens} focused on integrating it into food rescue systems. 
    Seamless integration within the food rescue platform is essential because it reduces the barriers for organizers to access and leverage \abr{RescueLens} for their decision-making; the easier it is to use \abr{RescueLens}, the more of an impact it will have. 
    As a simple example, we found that visually presenting these results is critical so organizers can better understand trends in an accessible manner. 
    \item \textbf{Metrics Beyond Accuracy} - We experimented with different configurations and underlying models during the development of  \abr{RescueLens}, where these settings varied along dimensions including precision, recall, and cost. 
    While larger models (e.g., GPT-4o) tend to perform better, we found that these models were significantly more expensive. 
    We engaged in conversations around tradeoffs between models, where we found that accuracy is not the sole arbiter. 
    We encourage future researchers to understand which metrics are most important to stakeholders rather than relying on a metric of convenience. 
    \item \textbf{Heterogeneity of Volunteer Feedback} - We built \abr{RescueLens} around volunteer feedback, and through the development of \abr{RescueLens}, we find that volunteer feedback is a rich source of information for a nonprofit. 
    Volunteer feedback in \abr{RescueLens} not only informs which donors and recipients require intervention but also details direction modifications and contact updates. 
    Such an idea holds across non-profits, as text-based feedback can reveal information about a non-profit's underlying health. 
    At the same time, human feedback is hard to parse due to inherent ambiguity. 
    We overcame this issue by developing clear standards when defining categories and using them to retrieve predictions. 
\end{enumerate}
\section{Discussion and Conclusion}
Food rescue organizations tackle food insecurity by redistributing excess food from donors to recipients via volunteers. 
Volunteer feedback is critical for a healthy relationship between non-profits and volunteers, yet manually processing feedback becomes cumbersome due to scale.
To tackle this problem, we develop \abr{RescueLens}, an LLM-based tool that automatically categorizes feedback at food rescue organizations and helps organizers take actions based on feedback. 
We deploy \abr{RescueLens} with our partners at 412 Food Rescue, and demonstrate that \abr{RescueLens} maintains high accuracy, precision, and recall.
Through a mixed methods study, we find that \abr{RescueLens} can help guide organizers at 412 Food Rescue identify which donors and recipients require intervention, thereby streamlining the feedback process. 
Throughout the process, we learned valuable lessons on how AI-based tools can be best integrated into non-profits. 
Future improvements to \abr{RescueLens} include a tracking system to measure which donors and recipients have been intervened upon and a system for answering volunteer questions based on prior feedback. 

Our results demonstrate the efficacy of LLMs in helping organizers better understand volunteer feedback.
Beyond food rescue, \abr{RescueLens} can be extended to other non-profits, and we hope our work outlines the steps needed for deployment. 
At the same time, non-profits vary in their access to data, types of feedback, and intervention priorities (e.g., how they intervene based on volunteer feedback).
\abr{RescueLens} is flexible enough to adapt based on non-profit specifics. 
For example, the donor and recipient scores can be modified based on non-profit priorities. 
To help with \abr{RescueLens} adoption, we include the code and prompts necessary for replication, which can help organizations with integration. 
Through the use of such a tool, non-profits can better quantify volunteer opinions and perspectives and improve non-profit health. 
\section*{Acknowledgements}
We thank Rex Chen for comments on an earlier draft of this paper. 
Co-author Raman is also supported by an NSF Fellowship. 
This work was supported in part by NSF grant IIS-2046640 (CAREER).

\appendix
\nobibliography*

\bibliography{references}
\section{Interview Questions}
\label{sec:questions_needfinding}
\subsection{Needfinding Study}
We include the questions from the needfinding study in Section~\ref{sec:user}: 
\begin{itemize}
\item What food rescue platform do you work at? 
\item What is your current role within the food rescue platform?
\item How is user feedback currently incorporated into the food rescue platform? 
\item What types of user feedback do you most commonly see in the system? 
\item What are the current pain points with the user feedback process? 
\item How do you manage issues related to donor pickup or recipient dropoff? 
\item If AI were to automatically analyze user feedback, how could that assist with the food rescue platform? 
\item How do you think \abr{RescueLens} could be incorporated into the food rescue workflow/pipeline? 
\item Are the predictions from \abr{RescueLens} useful?
\item Are there other categories of information that might be useful? 
\item Is it more important that our algorithm makes precise predictions or has a high recall rate? 
\item What features could we add to \abr{RescueLens} that would be helpful? 
\item Would this application be useful if deployed/implemented in practice? What barriers do you see to deployment? 
\item What components of the application would you remove?  
\item What limitations do you see with this application? 
\item How do you think the direction rewrite component can be incorporated into the food rescue workflow/pipeline? 
\item Are the rewritten directions useful?
\item What features could we add to the direction rewrite that would be helpful? 
\item Would the direction rewrite feature be useful if deployed/implemented in practice? What barriers do you see to deployment? 
\item What components of the direction rewrite would you remove?  
\item What limitations do you see with the direction rewrite component? 
\end{itemize}

\subsection{Deployment Study}
We include the questions from our deployment interview in Section~\ref{sec:deployment} below:
\begin{itemize}
    \item How often do you use \abr{RescueLens}?
    \item What happens after you look at \abr{RescueLens}? What fraction of the responses have an issue that leads to action? 
    \item Which labels are the most common or have the highest impact? 
    \item Do you have a sense for the precision or recall of \abr{RescueLens}
    \item Are there any types of feedback that would be good to label that we don't label right now? 
    \item Is it mostly one donor that pops up repeatedly, or is it more spread out? 
    \item What other features could we add in?
\end{itemize}

\section{Prompts}
\label{sec:prompts}
We include one prompt for predicting the inadequate food category and another prompt for direction rewrite (full prompts to predict all categories are in the attached code). 

\begin{WrappedVerbatim}
Description: As an analyst for our food rescue platform, your primary task is to carefully review feedback provided by volunteer drivers, with a specific focus on identifying issues caused by inadequate food provided by the donors.

Roles Explained:
The Donor - Provides the food.
The Volunteer Driver - Transports the food.
The Recipient - Receives the food.

Volunteer Feedback Analysis:
Volunteers interact with our app to claim tasks, collect food from donors, and deliver it to recipients. They leave comments and rate their experience post-rescue. Your task is to pinpoint and evaluate comments that reflect problems directly caused by inadequate food provision by the donor.

Guidelines for Analysis:
1. Inadequate Food: Assess whether the reported challenges or failures in the food rescue process were caused by inadequate food quantities provided by the donor. If the inadequate food is caused by issues with the donor (e.g. the donor forgot to respond), then do not mark it as inadequate food. 

Notes:
1. Consider comments as inadequate food issues when they indicate "no donation", "no pick-up", "meager food", or "limited access", "few", "nothing"...
2. Only consider comments as an inadequate food issue when recipient received no food, insufficient food, or had issues with the food itself. This DOES NOT include situations where a previous rescue trip picks up the food beforehand. 
3. Responses should be formatted in JSON to maintain uniformity and clarity across reports.

Example Comment Analysis:
1. For this rescue, the donor is Aldi Market; the recipient is North Long Beach Ministry Center. Comment: No donation today.

    "inadequate\_food": true, 
    "explanation": "The comment mentioned that there was no donation, which is a case of inadequate food."

2. For this rescue, the donor is Kroger; the recipient is Bethel Baptist Church. Comment: Nothing to donate. Everything they had put aside was burned.

    "inadequate\_food": true
    "explanation": "The comment mentioned that there was nothing to donate, a case of inadequate food."

3. For this rescue, the donor is 42-La Canada; the recipient is North Long Beach Ministry Center. Comment: I was told someone picked up earlier.

    "inadequate\_food": false
    "explanation": "The comment mentioned that someone came earlier, which explicitly SHOULD NOT be marked as inadequate food."

Now, it’s your turn. Analyze the following rescue
\end{WrappedVerbatim}

Similarly, for the direction rewrite, we have: 
\begin{WrappedVerbatim}
You are an analyst for a food rescue platform. Your ONLY job is to identify when volunteer feedback contains specific, actionable corrections to existing pickup or delivery directions.

**CRITICAL RULE: Only update directions when the volunteer explicitly states that existing directions are wrong, missing, or need specific changes.**
**CRITICAL RULE: Do not delete anything from the original directions, just add things on.**


## When to Update Directions (ONLY these cases):
- Volunteer says contact info is wrong/outdated
- Volunteer says the address/location is incorrect  
- Volunteer provides specific entrance/access corrections
- Volunteer states directions are missing key details

## When NOT to Update Directions:
- General complaints about the experience
- Issues with food quantity or quality
- Timing problems or delays
- Volunteer's personal difficulties
- Closed locations or unavailable contacts
- Anything that doesn't directly correct the written directions
- A lack of adequate directions

## Your Task:
Analyze the volunteer feedback and determine:

1. **Donor Direction Change**: true ONLY if volunteer explicitly corrects donor pickup directions
2. **Rewritten Donor Direction**: Only rewrite if change = true, using volunteer's specific correction. Only add on to the existing donor directions; DO NOT delete anything
3. **Recipient Direction Change**: true ONLY if volunteer explicitly corrects recipient delivery directions  
4. **Rewritten Recipient Direction**: Only rewrite if change = true, using volunteer's specific correction. Only add on to the existing recipient directions; DO NOT delete anything

## Output Format:
```json
{
  "donor_direction_change": boolean,
  "rewritten_donor_direction": "string (empty if no change)",
  "recipient_direction_change": boolean, 
  "rewritten_recipient_direction": "string (empty if no change)",
  "explanation": "Brief explanation of your decision"
}
```

## Key Examples:

**UPDATE NEEDED:**
- "The phone number goes to a gas station." → Update directions with the correct contact method
- "Go to Powell Street, not Alexander Street" → Update directions with the correct location
- "Use the side entrance, not the main door" → Update directions with the correct entrance

**NO UPDATE NEEDED:**
- "Terrible pickup experience" → General complaint, no direction correction
- "Too much food for my car" → Quantity issue, not direction issue  
- "Location was closed" → Timing issue, not direction correction
- "Nobody knew about the donation" → Communication issue, not direction correction

**Remember: You can only use information the volunteer explicitly provides. Do not assume or add details not stated in the feedback.**

## Examples:

### Example 1: Address Correction (UPDATE NEEDED)
**Input:**
```json
{
  "donor": "Production Facility",
  "recipient": "Powell Street",
  "volunteer_comment": "The map directions took me to Alexander Street for the pickup. Please adjust pick up location to Powell.",
  "donor_direction": "Enter the building with the big red door.",
  "recipient_direction": "Leave any food outside the steps."
}
```

**Output:**
```json
{
  "donor_direction_change": false,
  "rewritten_donor_direction": "",
  "recipient_direction_change": true,
  "rewritten_recipient_direction": "Leave any food outside the steps. Note: Go to Powell Street, not Alexander Street.",
  "explanation": "Volunteer explicitly corrected the pickup location from Alexander Street to Powell Street for the recipient."
}
```

### Example 2: Contact Info Correction (UPDATE NEEDED)
**Input:**
```json
{
  "donor": "2405 Southwest White Birch Drive",
  "recipient": "Ankeny Campus Bldg 6",
  "volunteer_comment": "Just fyi, the phone number for the contact at DMACC, building 6 (XXX?) was incorrect and directed to Casey's gas station.",
  "donor_direction": "",
  "recipient_direction": ""
}
```

**Output:**
```json
{
  "donor_direction_change": false,
  "rewritten_donor_direction": "",
  "recipient_direction_change": true,
  "rewritten_recipient_direction": "Contact phone number is incorrect - verify correct contact information for XXX at DMACC Building 6.",
  "explanation": "Volunteer explicitly stated the recipient contact phone number is wrong, requiring a direction update."
}
```

### Example 3: General Complaint (NO UPDATE)
**Input:**
```json
{
  "donor": "Food 4 Kids",
  "recipient": "Dumont House",
  "volunteer_comment": "Was told we may need to stop deliveries here because food isn't being used",
  "donor_direction": "Please call XXX when you are here.",
  "recipient_direction": ""
}
```

**Output:**
```json
{
  "donor_direction_change": false,
  "rewritten_donor_direction": "",
  "recipient_direction_change": false,
  "rewritten_recipient_direction": "",
  "explanation": "Comment about potentially stopping deliveries does not provide any corrections to existing pickup or delivery directions."
}
```

### Example 4: Experience Complaint (NO UPDATE)
**Input:**
```json
{
  "donor": "Pine Creek XXX",
  "recipient": "North Shore Senior Center",
  "volunteer_comment": "Terrible pickup! Nobody knew who 412 was. After 1/2 hour, I was given 3 boxes of apples. As I left, I was flagged down and given a cart full of leftover Easter candy. Arrived at drop location to find it was closed on Tuesday.",
  "donor_direction": "Please call the store prior to starting to confirm a 412 Food Rescue donation for the day. Ask for PIN# XXX.",
  "recipient_direction": "Please call Jackson once here."
}
```

**Output:**
```json
{
  "donor_direction_change": false,
  "rewritten_donor_direction": "",
  "recipient_direction_change": false,
  "rewritten_recipient_direction": "",
  "explanation": "Volunteer describes experience issues but provides no specific corrections to the existing pickup or delivery directions."
}
```

### Example 5: Food Quantity Issue (NO UPDATE)
**Input:**
```json
{
  "donor": "Kroger",
  "recipient": "North Shore Senior Center",
  "volunteer_comment": "I received more food than I could handle; my car could not fit in the many bags of bagels",
  "donor_direction": "Enter through the door down the stairs.",
  "recipient_direction": "Please call the store prior to starting to confirm donation."
}
```

**Output:**
```json
{
  "donor_direction_change": false,
  "rewritten_donor_direction": "",
  "recipient_direction_change": false,
  "rewritten_recipient_direction": "",
  "explanation": "Comment about food quantity does not provide any corrections to pickup or delivery directions."
}
```

### Example 6: Contact Difficulty (NO UPDATE)
**Input:**
```json
{
  "donor": "Kroger",
  "recipient": "North Shore Senior Center", 
  "volunteer_comment": "The pickup was awful; I tried contacting them but to no avail.",
  "donor_direction": "Enter through the door down the stairs.",
  "recipient_direction": "Please call Jackson once here."
}
```

**Output:**
```json
{
  "donor_direction_change": false,
  "rewritten_donor_direction": "",
  "recipient_direction_change": false,
  "rewritten_recipient_direction": "",
  "explanation": "General complaint about contact difficulty does not specify what corrections are needed to the existing directions."
}
```

---

**Now analyze this feedback:**
\end{WrappedVerbatim}

\section{Direction Rewrite Rubric}
\label{sec:rubric}
We assess rewritten directions according to the following rubric:
\begin{enumerate}
    \item \textbf{Helpfulness} - Is the direction rewrite helpful in assisting the volunteers?
    \begin{itemize}
        \item 5 - The direction is very helpful and provides essential information for the volunteers to complete the trip
        \item 4 - The direction is somewhat helpful; completing the trip is a bit easier, though the volunteers would do just fine without the directions
        \item 3 - The direction is neither helpful nor hurtful; completing the trip would be just as easy for the volunteers without the directions
        \item 2 - The directions are somewhat hurtful; completing the trip is a bit harder, though the volunteers would not be much worse off with the directions
        \item 1 - The directions are significantly harmful; completing the trip is much harder when using the directions
    \end{itemize}
    \item \textbf{Novelty} - Does the direction include significantly new information? 
    \begin{itemize}
        \item 5 - The directions include significantly new information not present in the original directions. It could also place a note for new information to be included.
        \item 4 - The directions include some new information not present in the original directions
        \item 3 - The directions are mostly the same as the information present in the original directions, maybe with some slight rewording. 
        \item 2 - The directions either leave out some important information from the original directions, or include some made-up information
        \item 1 - The directions are almost entirely made up, or are a complete erasure of the previous directions
    \end{itemize}
    \item \textbf{Clarity} - Are the rewritten directions clear? 
    \begin{itemize}
        \item 5 - The new directions are very clear and understandable
        \item 4 - The new directions can be understood fairly quickly, though it takes a bit of effort
        \item 3 - The new directions are somewhat understandable; one or two phrases might be unclear
        \item 2 - The new directions are fairly hard to understand; many phrases need to be reworded or re-read for understanding. 
        \item 1 - The new directions are completely illegible 
    \end{itemize}
\end{enumerate}

\end{document}